\documentclass[aps,prb,reprint,superscriptaddress,longbibliography,floatfix]{revtex4-1}

\usepackage{amsmath}
\usepackage{amssymb}
\usepackage{graphicx}

\usepackage[english]{babel}
\usepackage{braket}
\usepackage{breakcites}
\usepackage{bm,braket}
\usepackage{float}
\usepackage{graphicx}
\usepackage[caption=false]{subfig}
\usepackage{amsfonts}
\usepackage{amsmath}
\usepackage{amssymb}
\usepackage{color}

\usepackage[usenames,dvipsnames]{xcolor}

\usepackage[utf8]{inputenc}
\usepackage[T1]{fontenc}

\usepackage[colorlinks]{hyperref}
\definecolor{winered}{rgb}{0.5,0,0}
\hypersetup
{
colorlinks=true,
linkcolor=winered,
urlcolor={winered},
filecolor={winered},
citecolor={winered},
allcolors={winered}
}

\usepackage[markup=default]{changes}

\graphicspath{{./figs/}}

\usepackage{letltxmacro}

\LetLtxMacro{\ORIGselectlanguage}{\selectlanguage}
\makeatletter
\DeclareRobustCommand{\selectlanguage}[1]{%
  \@ifundefined{alias@\string#1}
    {\ORIGselectlanguage{#1}}
    {\begingroup\edef\x{\endgroup
       \noexpand\ORIGselectlanguage{\@nameuse{alias@#1}}}\x}%
}
\newcommand{\definelanguagealias}[2]{%
  \@namedef{alias@#1}{#2}%
}
\makeatother

\definelanguagealias{en}{english}
\definelanguagealias{EN}{english}

\usepackage{xargs}
\newcommandx{\greencom}[2][1=]
{\todo[inline, color=green!40,#1]{#2}}
\newcommandx{\bluecom}[2][1=]
{\todo[inline, color=blue!40,#1]{#2}}

\begin{document}
\sloppy
\emergencystretch 3em

\title{Resonant Raman scattering of single molecules under
strong cavity coupling and ultrastrong optomechanical coupling in plasmonic resonators:  phonon-dressed polaritons}

\author{Stephen Hughes}
\affiliation{Department of Physics, Engineering Physics and Astronomy,
Queen's University, Kingston, ON K7L 3N6, Canada}

\author{Alessio Settineri}
\affiliation{Dipartimento di Scienze Matematiche e Informatiche,
Scienze Fisiche e Scienze della Terra, Universit`a di Messina, I-98166 Messina, Italy}

\author{Salvatore Savasta}
\affiliation{Dipartimento di Scienze Matematiche e Informatiche,
Scienze Fisiche e Scienze della Terra, Universit`a di Messina, 
I-98166 Messina, Italy}

\author{Franco Nori}
\affiliation{Theoretical Quantum Physics Laboratory, RIKEN Cluster for Pioneering Research, Wako-shi, Saitama 351-0198, Japan}
\affiliation{Physics Department, The University of Michigan, Ann Arbor, Michigan 48109-1040, USA}

\begin{abstract}
Plasmonic dimer cavities can induce
extreme electric-field hot spots that allow one to access
ultrastrong coupling regimes
using Raman-type spectroscopy on single vibrating molecules. Using a 
generalized master equation, 
we study resonant Raman scattering
in the strong coupling regime of cavity-QED, when also in the  vibrational ultrastrong coupling regime, leading to ``phonon-dressed polaritons''.
The master equation rigorously includes 
spectral 
baths for
the cavity and vibrational degrees of freedom, as well as  a pure dephasing bath for the resonant two-level system, which  play a significant role. 
Employing realistic parameters  for gold dimer
cavity modes,  we investigate the
 emission spectra in  several characteristic strong-coupling regimes,
leading to extremely rich spectral resonances
due to an interplay of phonon-modified polariton states and bath-induced resonances. 
We also show explicitly the failure of the standard master equation in these quantum nonlinear regimes.
\end{abstract}
\maketitle

{\section{Introduction}}

The celebrated Raman effect is based on the inelastic scattering of monochromatic incident
radiation, whereby 
coherent optical fields couple to molecular vibrations and scatter at phonon-shifted frequencies with respect to the excitation frequency~\cite{Raman}.
 The usual Raman interaction leads to
 Stokes emission (lower energies) and anti-Stokes
 emission (higher energies).
 Although most Raman experiments 
 deal with very weak scattering cross sections,  surface-enhanced Raman spectroscopy (SERS) with metal nanoparticles (MNPs) can boost this interaction by many orders of magnitude. 
 
 While there is much interest in using MNPs to explore new regimes in quantum plasmonics~\cite{PhysRevB.84.045310,TameQuantumPlasmonics,doi:10.1117/1.JNP.10.033509,doi:10.1021/acs.nanolett.7b02222,StrongCoupling,PhysRevLett.105.263601,Ge2013,ControllingSE,LargePF,PhysRevB.92.205420,PhysRevA.101.013814},
a major problem  for enhancing quantum light-matter interactions is  metallic losses. In contrast to dielectric cavity systems, the quality factors for MNPs are only around
$Q \sim 5-20$, resulting in
significant cavity decay rates, $\kappa$,  typically to large to resolve  the higher lying quantum states.
However, plasmonic resonator cavity modes also yield extremely small mode volumes that can  compensate for the large losses. Thus, treating the plasmonic system as a ``bath''
is not necessarily a good approximation, and quantum correlations 
can  be important.

In the field of optomechanics,
photon-photon interactions with vibrational interactions can give rise
to interesting correlation effects~\cite{Rabl},
but accessing such regimes 
has proven experimentally difficult, and often the problem is treated 
through various linearization procedures, where the photons are not entangled with the phonons.
Recently, it has also been shown how
SERS  can be viewed as an effective enhancement of the optomechanical coupling between the localized surface plasmon resonance and the vibrational mode of the molecule, leading to new ideas in {\it molecular optomechanics}~\cite{Roelli,Schmidt,KamandarACS}.
In such regimes, 
the plasmon mode and vibrating molecule mimic the
 optomechanical coupling scenario with a 1D cavity mode in the presence of a vibrating mirror~\cite{Roelli}.
Potential advantages of molecular optomechanics, over dielectric based systems, include access to higher vibrational frequencies and larger optomechanical coupling rates.
 Several key experiments in this area have emerged,
including 
pulsed molecular optomechanics \cite{PulsedOptomechanics}. 
For hybrid metal-dielectric resonances,
molecular optomechanics in
 the sideband-resolved regime has also been predicted~\cite{Dezfouli2019}.
  Vibronic strong coupling effects in Raman scattering have also been studied~\cite{delPino2015}, 
using a quantum theory of 
strong coupling of collective molecular vibrations (that are IR active) within a microcavity.
 
\begin{figure}[t]
\includegraphics[width=0.99\columnwidth]{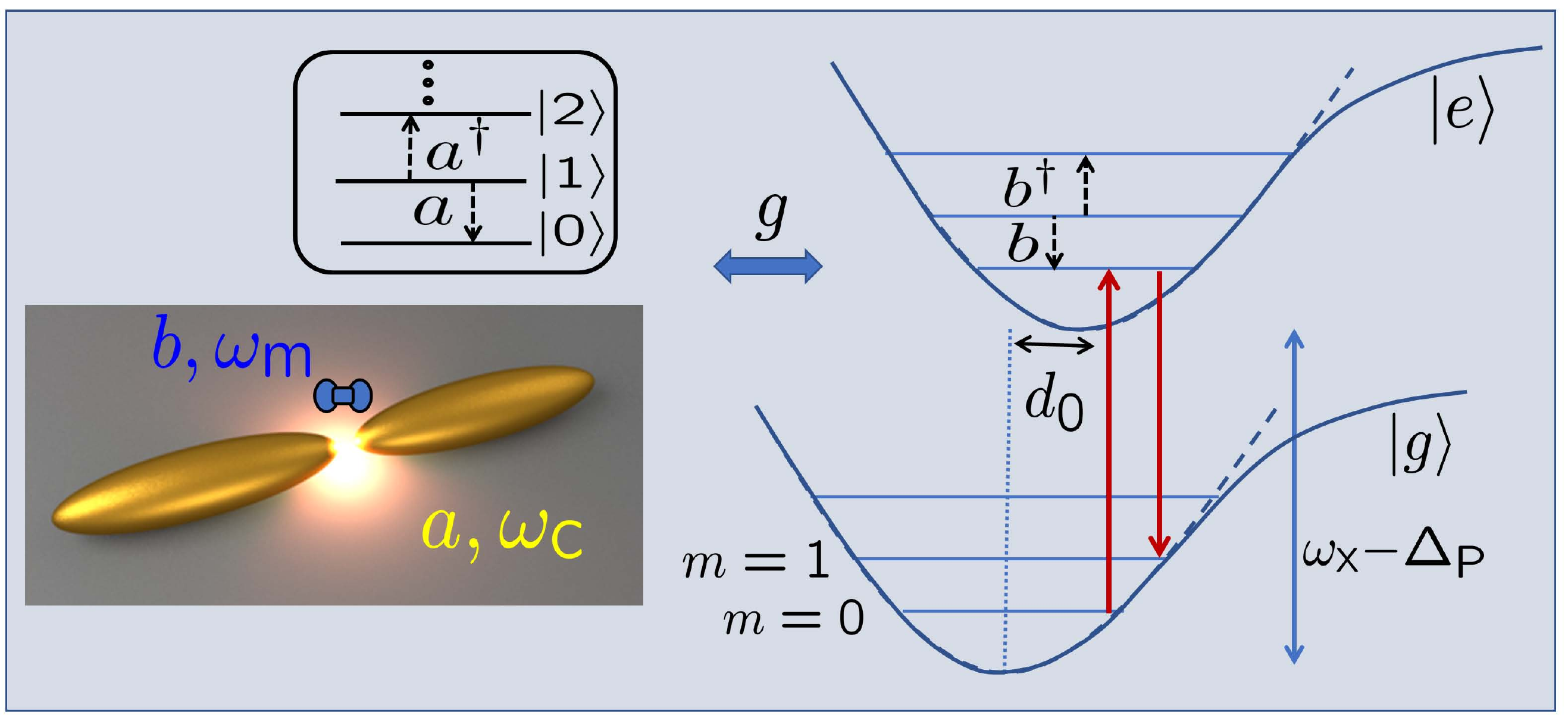}
\caption{Schematic of a plasmonic field hot-spot coupled
to a vibrating molecule ($b,b^\dagger$) and the cavity mode ladder states
($a,a^\dagger$). At the right, we also show the  two  electronic manifolds, each containing
a subset of phonon levels ($k=0,1, \cdots$) separated by
$\omega_{m}$, and coupled to the cavity mode through $g$; the  excited manifold is shifted
down from the bare exciton resonance ($\omega_{x}$) through the polaron shift,  $\Delta_{P}=d_{0}^2\omega_{ m}$,
and shifted by the normalized displacement  $d_0$. 
 \label{fig1}}
\end{figure}

One way to boost the optomechanical rates
 even further is to use {\it resonant} electronic excitations,
 which  brings in the traditional domains of cavity QED (see Fig.~\ref{fig1}), and modified optomechanical coupling with Fermionic statistics.
Molecular two-level system (TLSs) coupled to MNPs have already shown experimental signatures of vacuum Rabi splitting~\cite{StrongCoupling}, where it is also noted there there are rich Raman transitions involved~\cite{StrongCoupling}. 
To understand such emerging systems,
one must couple the physics of resonant SERS
and traditional cavity-QED  physics, 
while properly accounting for vibrational ultrastrong coupling (USC) and
complex bath interactions,
in regimes 
where traditional master equations can significantly fail.

Several resonant
SERS schemes have been theoretically studied with 
various limitations.
 A bad cavity limit has been used to explore 
the interplay between 
TLS driving 
and vibrational coupling~\cite{PhysRevA.100.043422}, where the Mollow triplet resonances can be split by phonon interactions; the theory employed a standard master equation (SME) with simple TLS pure dephasing processes.
Also using a SME, Ref.~\onlinecite{Neuman2020}
 studied resonant SERS in the good cavity limit, presenting useful analytical solutions and numerical results of hybrid resonances; however, for vibrational USC, the SME breaks down.
Recently,
USC in molecular cavity-QED
using the quantum Rabi model  
has been explored~\cite{Calvo2020},
where  
model Hamiltonians and dynamical coupling in a one photon  subspace were studied, though dissipation was neglected.


In this work, we present a 
theory  of  MNP-based single molecular optomechanics
in the strong cavity-coupled {{\it nonlinear}} resonant Raman regime, fully accounting
for the dynamics of the system level operators
for the TLS, cavity mode, and a molecular phonon mode.
We  rigorously treat system-bath dissipation 
by using a generalized master equation (GME) approach,
modelled with characteristic spectral functions for the photon and phonon baths as well as excitonic pure dephasing. 
We also show explicitly that the SME fails in such regimes.
Indeed, the GME is shown 
to yield much richer spectral features because of bath-induced interactions.
We explore several regimes of simultaneous strong cavity-exciton coupling and vibrational coupling in the USC regime~\cite{USCR1,RevModPhys.91.025005}, showing
rich cavity emitted spectra, including phonon dressed strong coupling with resonant Raman scattering. These effects are also influenced by the presence of multiple bath-induced asymmetries and modified resonances.\\

\section{Theory}

\label{sec2}

To make a connection with off-resonance Raman interactions,
it is useful to first consider the  standard optomechanical Hamiltonian~\cite{RevModPhys.86.1391} ($\hbar=1$), 
\begin{equation}
H_{s}  =\omega_{c}\,a^{\dagger}a+\omega_{ m}\,b^{\dagger}b- g_{\rm om}\,a^{\dagger}a\left(b^{\dagger}+b\right) ,\label{eq:H10}
\end{equation}
where $\omega_{m}$ is the molecular vibrational mode frequency, 
$\omega_{c}$  cavity mode resonance frequency,
and $a,a^\dagger$ and $b,b^\dagger$
are the annihilation and creation operators for the cavity mode and phonon mode, respectively. 
The eigenenergies of 
Eq.~\eqref{eq:H10}
are~\cite{agarwal_quantum_2012}
\begin{equation}
{\omega}_{n,k}= n\omega_{ c} +  k \omega_{m}
-n^2 { g_{\rm om}^2}/{\omega_{ m}},
\end{equation}
with the corresponding
eigenstates,
\begin{equation}
\ket{\Psi_{n,k}}{=}
{D}^\dagger\left ({g_{\rm om}n}/{\omega_{ m}}\right)
\ket{n,k},
\end{equation}
where ${D}$ is the displacement operator. 
The displacement of phonons
also depends on the 
 on the number of photons.
 This results in photon manifolds that contain phonon sub-levels, and each manifold is shifted down in energy
 from the cavity resonance. 
In a polaronic frame, the cavity mode
 frequency changes from $\omega_c \rightarrow \omega_{c}-g_{\rm om}^2/\omega_{\rm m}=\omega_{c}-d_0^2 \omega_{m}$,
 where $d_0$ is a dimensionless displacement~\cite{PhysRevA.100.043422}.

{For resonant Raman interactions, we assume that the 
plasmonic MNP interacts with electronic TLS vibrational degrees of freedom
through a  Huang-Rhys theory, first formulated for F centers in 1950 \mbox{\cite{1950,Kubo1955,MahanBook,PhysRevA.100.043422}}, which has been widely used
to model vibronic interaction in molecules~\mbox{\cite{delPino2015,Neuman2018}},
and has the same form as electron-phonon scattering in  quantum dot systems~\mbox{\cite{PhysRevB.86.241304,PhysRevB.92.205406,PhysRevB.90.035312,PhysRevB.65.235311,MahanBook}}}
The
system Hamiltonian is
\begin{align}
H_{ s}  &= \omega_{ c} a^{\dagger}a+\omega_{ x}\,\sigma^+\sigma^- +
\omega_{ m}b^\dagger b + d_0\omega_{ m} \sigma^+\sigma^-  (b^{\dagger}+b)  
\nonumber \\
&+ g(\sigma^+ a +  a^\dagger \sigma^-),
\label{eq:H2}
\end{align}
where $\omega_{ x}$ is the exciton resonance
frequency, 
and the latter term is the Jaynes-Cummings interaction,
where we assume $g \ll \omega_c$,
so as to neglect USC effects from the cavity-TLS interaction. However, USC effects from  vibrational coupling are fully included.

With system-bath interactions included,  the SME for resonant SERS with a single MNP cavity mode is~\cite{Breuer,Carmichael}
\begin{align}
\frac{d\rho(t)}{dt} & =-{i}\left[H_{ s},\rho(t)\right] + \frac{\kappa}{2}\,\mathcal{D}[a]\rho(t)
+\frac{\gamma_\phi}{2}\,\mathcal{D}[\sigma^+\sigma^-]\rho(t) \nonumber\\
\nonumber\\
 & +\frac{\gamma_{m}\left(1+\bar{n}_m\right)}{2}\,\mathcal{D}[b]\rho(t) + \frac{\gamma_{m}\bar{n}_m}{2}\,\mathcal{D}[b^{\dagger}]\rho(t),
\label{eq:rho}
\end{align}
where $\kappa$ is the cavity decay rate (which dominates over spontaneous emission),
$\gamma_\phi$ is a possible pure dephasing rate 
of the TLS,
$\gamma_{ m}$ is the vibrational decay rate, and
$
\bar{n}_{ m}={1}/({{\rm exp}\left(\omega_{m}/T\right){-}1}),
$
is the thermal population of the vibrational mode at temperature $T$ (in units of $k_B=1$). 
The system Hamiltonian is given in
Eq.~\eqref{eq:H2}, and
the Lindblad superoperator 
is 
defined by
\begin{align}
    \mathcal{D}[O]\rho(t){=}O\rho(t) O^{\dagger}{-}0.5O^{\dagger}O\rho(t){-}0.5\rho(t) O^{\dagger}O.
\end{align}

A known problem with the SME
is that it neglects the internal coupling
between the system operators when deriving
the system-bath interactions~\cite{Carmichael1973,Ge2013,Beaudoin2011,Settineri2018}.
To address this problem,
we exploit a GME approach~\cite{Settineri2018},
which takes into account the dressed-states' coupling to the bath reservoirs: 
\begin{equation}
\label{me}
\frac{{d}}{{dt}}\rho = -{i}[H_{ s},\rho] 
+ \mathcal{L}_{{G}}\rho
+ \mathcal{L}_{{G}}^\phi\rho
,
\end{equation}
where 
the dissipator term 
\begin{align}
\label{gme1}
&\mathcal{L}_{{G}} \rho = \frac{1}{2}
\sum_{\alpha={ c},{ m}}\, \sum\limits_{\omega,\omega'>0} \nonumber \\ &\ \ \ \ \Gamma_\alpha(\omega)(1+\bar{n}_\alpha(\omega))[x^+(\omega)\rho x^-(\omega') - x^-(\omega')x^+(\omega)\rho] \nonumber \\
&+ \Gamma_\alpha(\omega')(1+\bar{n}_\alpha(\omega'))[x^+(\omega)\rho x^-(\omega') - \rho x^-(\omega')x^+(\omega)] \nonumber \\
& + \Gamma_\alpha(\omega)\bar{n}_\alpha(\omega)[x^-(\omega')\rho x^+(\omega) - \rho x^+(\omega)x^-(\omega')] \nonumber \\
 & + \Gamma_\alpha(\omega')\bar{n}_\alpha(\omega')[x^-(\omega')\rho x^+(\omega) - x^+(\omega)x^-(\omega')\rho]  \nonumber \\
&+ \Gamma_\alpha'(T) [2 x_\alpha^0 \rho x_\alpha^0 - x_\alpha^0 x_\alpha^0 \rho - \rho x_\alpha^0 x_\alpha^0 ],
    \end{align}
     describes emission ($\Gamma(1+\bar n)$ terms), incoherent excitation ($\Gamma \bar n$ terms),
    and pure dephasing ($\Gamma'$ terms),
    from the cavity and phonon baths.
Note that, importantly, we do not
make any secular approximations, though  neglect counter-rotating wave terms  that oscillate at $\pm(\omega+\omega')$.
 The dressed-state operators,
    solved in a basis of energy eigenstates with respect to $H_{ s}$,
    are defined through
       \begin{align}
    x^+_{\alpha}(\omega)&=\sum_{j,k>j} \braket{j|(O_\alpha)|k}
    \ket{j}\bra{k}
\nonumber \\
  &x^0_{\alpha} 
  =\sum_{j}
  \braket{j|(O_\alpha)|j}  \ket{j}\bra{j},  
    \end{align}
    where $\omega=\omega_k-\omega_j>0$,
     $x^-=(x^+)^\dagger,$
     with 
     $O_c=a+a^\dagger$ and $O_{ m}=b+b^\dagger$.

    We have assumed Ohmic bath functions for both
    cavity and phonon baths [$J_\alpha(\omega)=\Gamma_\alpha \omega/2\pi\omega_\alpha$], and the
decay rates are then defined from
$
        \Gamma_\alpha(\omega) = {\gamma_\alpha\omega}/{\omega_\alpha}$,
    and
     $
        \Gamma'_\alpha(T) =
        {\gamma_\alpha T}/{\omega_\alpha}$ (bath-induced
    pure dephasing); the vibrational pure dephasing term becomes especially important at elevated temperatures.
For the TLS pure dephasing  term
$\mathcal{L}_{{G}}^\phi\rho$ (with $O_{x}=\sigma^+\sigma^-$),
we use a common spectral function for molecules~\cite{delPino2015}, 
   $J_{ \phi}(\omega)
    =\eta \omega e^{ -{\omega^2}/{\omega_{\rm cut}^2}}$,
where $\omega_{\rm cut}=160\,$meV is the cut-off frequency, and
$\eta$ is the coupling strength
which we define from 
$\gamma_\phi=\Gamma_\phi(0)=10$ meV at room temperature, and scales linearly with temperature.
We can also define the pure dephasing rates
explicitly, including
$\bar n_x(\omega)$ (upwards transition) and $1+\bar n_x(\omega)$ (downward transition):
\begin{align}
    \Gamma^{\downarrow}_\phi(\omega)
    &= 2\pi J_\phi(\omega) (1+\bar n_x (\omega)),
    \ \ \omega \geq 0 , \nonumber \\
    \Gamma^{\uparrow}_\phi(\omega)
    &= 2\pi J_\phi(-\omega) \bar n_x (-\omega), \ \ \omega<0.
\end{align}


 
 Finally,  in the interaction picture at the laser frequency
 $\omega_L$, we also add in a coherent cavity pump term,
%
$H_{\rm pump}
      = \Omega(a+a^\dagger)$,
  where $\Omega$ is the continuous wave (CW) Rabi frequency. This term also transforms
to the dressed-state basis, so that
     $H_{\rm pump}
     \rightarrow \Omega(x_{ c}^+ + x_{ c}^-)$,
which is included  after diagonalizing the density matrix
 from the solution of $H_{ s}$.
We note that while previous studies with the GME
have focused on the quantum Rabi model and optomechanical interactions~\cite{Settineri2018}, separately; here
we have a combined interaction with three system operators, and we have also explicitly accounted for a TLS pure dephasing bath as well as a coherent pump field
for the cavity mode. This approach also enables us to have clear insights into 
the underlying dressed resonances.
 
 Numerically, we obtain
 an appropriately large number of energy states from a basic
 of $n$ photons, $m$ phonons and the
 TLS, and truncate to the lowest 
 $N$ levels in the dressed-state basis, and check that this truncation is numerically conserved for each problem studied below. 
 Simulations are performed using QuTiP~\cite{qutip}.


\section{Results}

\subsection{System parameters}

For the MNP of interest, we  consider
cavity mode parameters representative of
metal dimers with a small gap,
ranging from $0.5{-}2$\,nm  to create a pronounced field hot-spot~\cite{Dezfouli2019}. 
The dissipative open-cavity modes can be quantitatively
described using  quasinormal modes (QNMs)~\cite{PhilipACS,hybrid,lalanne_light_2018}, which are solutions to the 
source-free Maxwell equations with open boundary conditions.
Even for the smallest gap, the entire response is very well explained with a single 
QNM~\cite{Dezfouli2019}.
The QNM complex eigenfrequencies are
defined from~$\tilde \omega_c = \omega_c-i\gamma_c$,
where $\kappa=2\gamma_{c}$, $Q=\omega_{ c}/\kappa$ and the effective mode volume is obtained from the normalized QNM spatial profile
at the dimer gap center~\cite{PhilipACS}.
For resonant SERS, we 
choose values of $d_0=0.2-1$~\cite{PhysRevA.100.043422,SnchezCarrera2010},
where $d_0=1$ corresponds to $g_{\rm om}=\omega_{ m}$.
For the molecular vibrational mode, we consider 
a smaller frequency oscillation at $\hbar\omega_{ m}=20\,{\rm meV}$
as well as a higher  frequency oscillation at $\hbar\omega_{ m}=160\,{\rm meV}$,  
with $\gamma_{ m}=0.8~$meV.
For the cavity mode, we use
$\kappa=100\,$meV,
and $\omega_{c}=\omega_{ x} = 1.7~$eV.

\subsection{Strong cavity-exciton coupling and ultrastrong vibrational coupling with $g>\omega_{ m}$}

We first consider a regime
where $\omega_{ m}= 20\,$meV
and $g=\omega_{ m}= \kappa$,
with $d_0=0.2$.
\begin{figure}[hb]
\includegraphics[width=0.99\columnwidth]{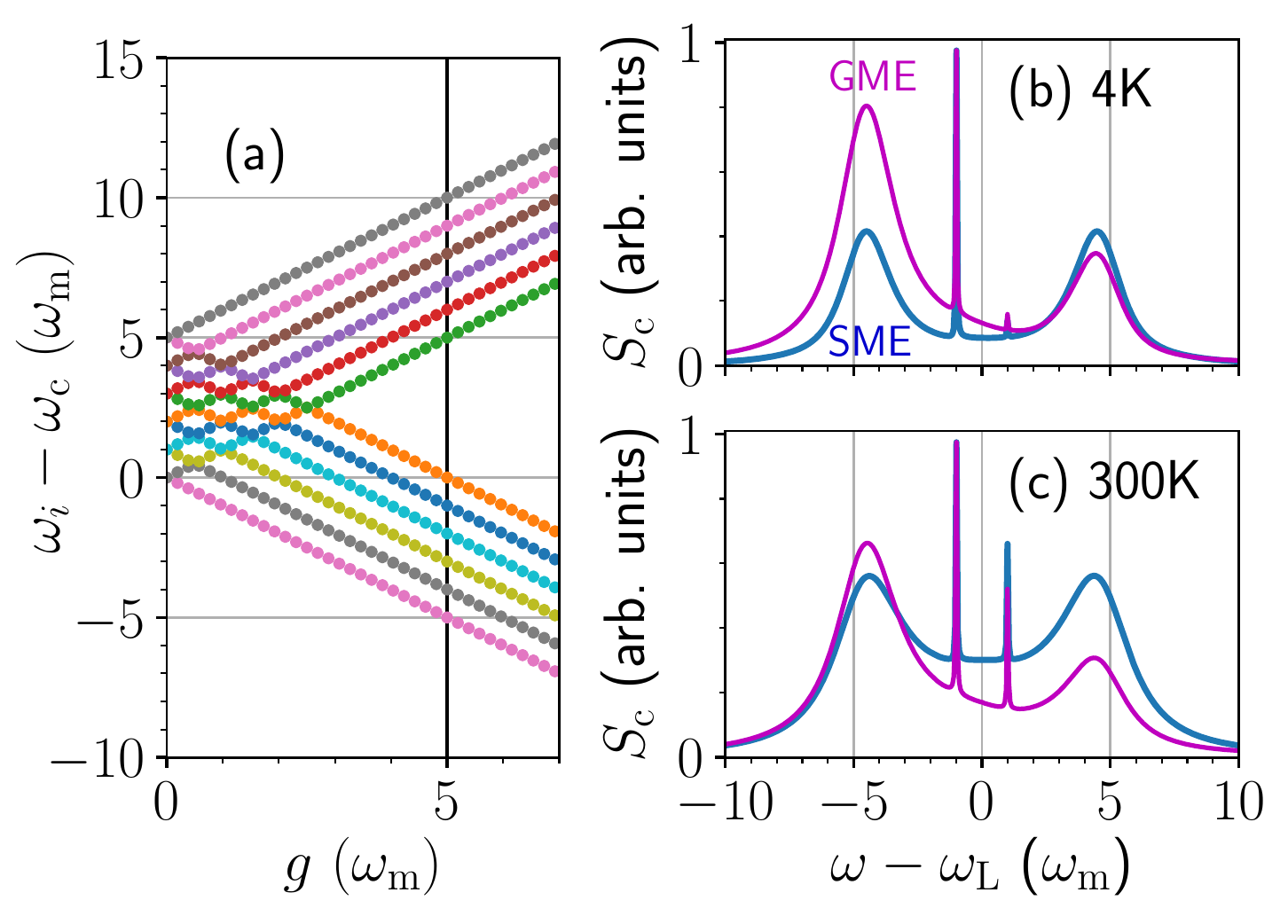}
\vspace{-0.5cm}
\caption{
(a) Eigenfrequencies of the system Hamiltonian
(Eq.~\eqref{eq:H2}) for  the $n=1$ photon manifold as a function of $g$ with $d_0=0.2$.
(b) Cavity emitted spectra with a 
coherent drive
($\omega_{ L}=\omega_{ x}=\omega_{ c}$)
at $g=5\omega_{ m}$ and $4~$K, showing the SME solution [Eq.~\eqref{eq:rho}, blue solid curve] and GME solution
[Eq.~\eqref{me}, magenta solid curve].
(c) Cavity emitted spectra at $300~$K.
 \label{fig:2}}
\end{figure}

Figure~\ref{fig:2}(a) shows the 
eigenenergies
without dissipation as a function of $g$, for
the first 6 phonon levels
in the $n=1$ photon subspace. At 
$g=5\omega_{ m}$, the
$k=1$ lower polariton shifts down
to $-5\omega_{ m}$, with the phonon states
split by exactly one phonon frequency; a similar trend happens at the upper polariton states. In the $n=0$ photon subspace, we simply obtain constant energy levels split by $\omega_{ m}$. Since photon transitions can take place from any of these excited phonon states, in general the splitting will always
be less than $\omega_{ m}$, even when
$\kappa \ll g$.

To access these energy states in the presence of resonance Raman scattering, we calculate the emitted spectrum
in the presence
of the CW field $\Omega=0.25g$. We calculate the cavity emitted spectrum of the hybrid system from
$
S_{ c}(\omega)  \equiv {\rm Re}\{ \int_{0}^{\infty}dt\,e^{i(\omega_{L}-\omega)t}
[\langle x_{ c}^-\left(t\right) x_{ c}^+(0)\rangle _{{\rm ss}} -\langle x_{ c}^{-}\rangle_{{\rm ss}}\langle x_{ c}^+\rangle _{{\rm ss}}] \}$,
where the expectation values are taken over the system steady state.

Figure~\ref{fig:2}(b) displays the results at
a  temperature of
$4\,$K, showing the first Stokes and anti-Stokes sidebands as $\pm \omega_{ m}$, and
polariton peaks around $\pm 4.5 \omega_{ m}$; these resonances are not increased with smaller $\kappa$ and are due to the collective summation of the phonon dressed states contributing to the emission linewidth.
The sharp Raman spectral features 
stem from long-lived Raman oscillations that
damp as the bath-modified
phonon decay rates, and these damp further with increasing drive strengths.
Interestingly, we also see a pronounced
asymmetry in the GME calculations~\cite{Cao2011}, which for this example is 
mainly coming from the spectral properties of the
TLS dephasing bath. This shows that even though the
zero phonon line has negligible dephasing, the properties of the bath at the dressed resonances has a substantial influence on the oscillator strengths. The bath-induced resonances cause downward transitions between the upper polariton and lower polariton states, while the transitions from the lower to higher states are negligible at low temperatures. 

Figure~\ref{fig:2}(c) shows the results at 
$T=300$~K, which confirms that these effects survive at elevated temperatures, and the bath-induced asymmetry is still visible but less pronounced in the GME, since now there is a larger probability also for bath-induced upwards transitions between the
polariton states; the anti-Stokes Raman transition is also much more visible.

A spectral  asymmetry has been shown also using a SME approach with effective Lindblad operators in a single exciton subspace~\cite{Neuman2020},
though the  general spectra trends as a function of temperature are quite different. For example, we obtain clear Stokes and anti-Stokes signals even at
300~K, and  qualitatively different spectra at low temperatures.
Apart from requiring multiple input states for our GME model,
it is important to note that 
for a bath in thermal equilibrium, 
one must satisfy
 detailed balance~\cite{RevModPhys.82.1155,Beaudoin2011},
$\Gamma^\uparrow_\phi(\omega)= e^{\omega/T} \Gamma^\downarrow_\phi(\omega)$.
Outside the regime of resonant Raman scattering and nonlinear driving,
the role of phonon-induced asymmetry in vacuum Rabi spitting has  been shown to be important in various cavity-QED systems, including 
 molecules~\cite{delPino2015,Neuman2018}
and electron-phonon scattering in quantum dot systems~\cite{PhysRevB.65.235311,PhysRevB.78.035330,PhysRevB.92.205406,PhysRevB.90.035312,PhysRevB.102.235303}.

\subsection{Strong cavity coupling and ultrastrong vibrational coupling with $\omega_{ m}>g$}

We next explore
polariton-dressed Raman transitions, and 
consider a regime where $\omega_{ m}= 160\,$meV
and $g= 2/3\, \omega_{\rm m}$,
for several different values of 
$d_0$, and all other parameters remain the same as before. 
%
\begin{figure}[ht]
\includegraphics[width=0.99\columnwidth]{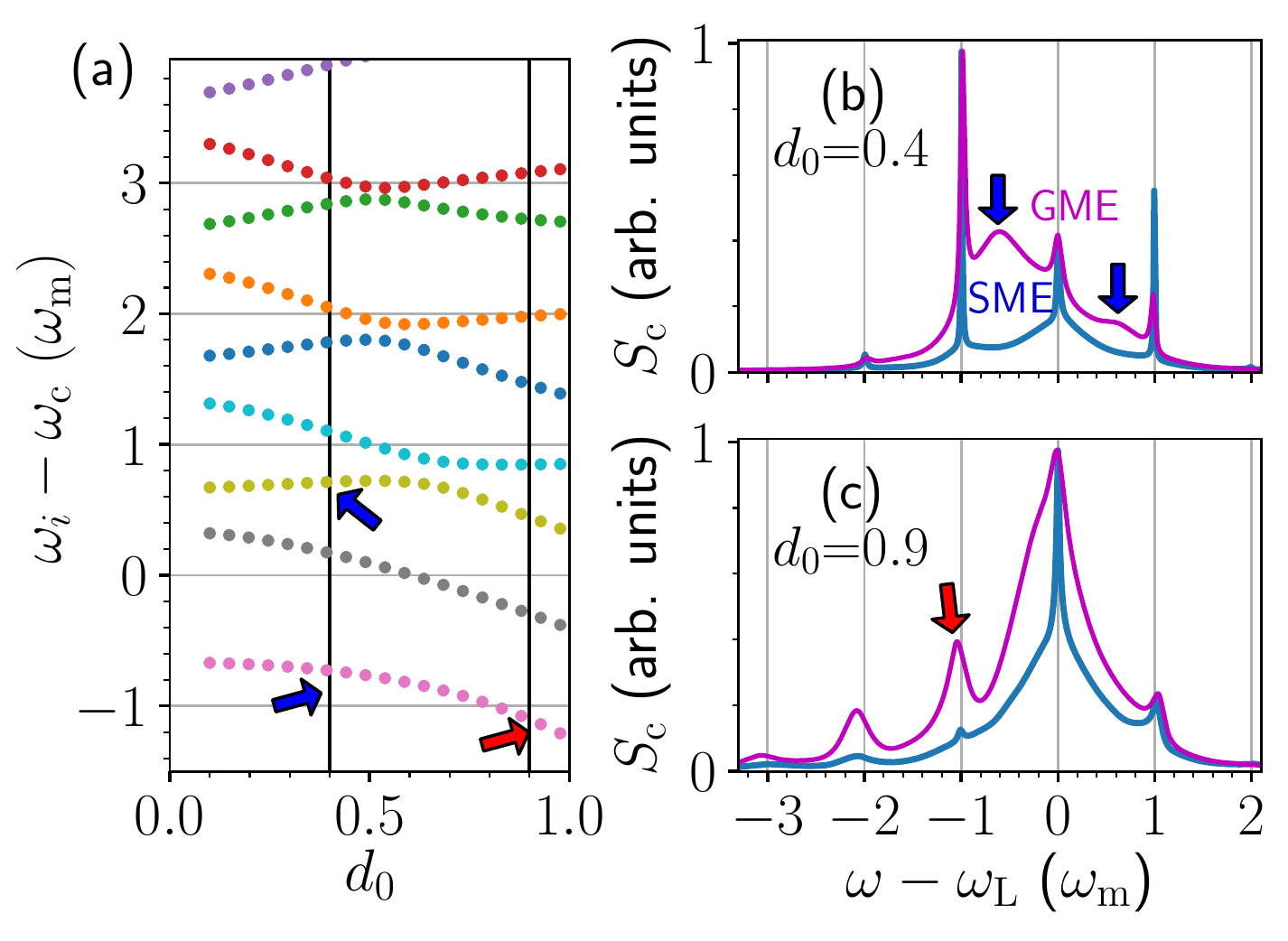}
\vspace{-0.5cm}
\caption{
(a) Eigenfrequencies of the system Hamiltonian
(Eq.~\eqref{eq:H2}) for  the $n=1$ photon manifold as a function of $d_0$ with $g=80~$meV and
$\omega_{ m}=120~$meV.
(b) Cavity emitted spectra with a coherent drive
at $d_0=0.4$, at $T=4$~K, showing the SME [Eq.~\eqref{eq:rho}, blue solid curve] and GME solution
[Eq.~\eqref{me}, magenta solid curve].
(c) Cavity emitted spectra at $d_0=0.9$.
Arrows show some characteristic resonances.
 \label{fig:3}}
\end{figure}

Figure~\ref{fig:3}(a) shows the system eigenenergies
as a function of $d_0$,
which are much richer than the previous example. The
first and third levels correspond to the polariton states with $k=0$ (zero phonons), which can be dressed by an increasing
$d_0$, and the third level anticrosses with
the $k=2$ lower polariton state around
$d_0=0.6$. The second resonance (from the bottom) corresponds to the $k=1$ lower polariton state which decreases as a function of $d_0$.
Figure~\ref{fig:3}(b) shows the cavity emitted spectra
at $d_0=0.4$, and we label some of the zero-phonon polariton states with arrows; these resonances clearly shows up before the first Raman sidebands and interestingly are not visible in the SME simulations; they also become more pronounced for smaller $\kappa$.
Figure~\ref{fig:3}(c) shows the  cavity emitted spectra at $d_0=0.9$, where we see that the first Stokes resonance is shifted down
from $-\omega_{ m}$; there is also rich structure at the red side of the center frequency, which originates from the second lowest resonance in (a). These results are qualitatively similar at elevated temperatures, with the SME drastically failing throughout the entire frequency range.
In this example, asymmetries form the TLS pure dephasing bath
are negligible, but are mainly caused by the larger $d_0$ causing phonon-modifications to the  vibrational and cavity emission and excitation dissipator terms in the GME.

 \begin{figure}[th]
\includegraphics[width=0.99\columnwidth]{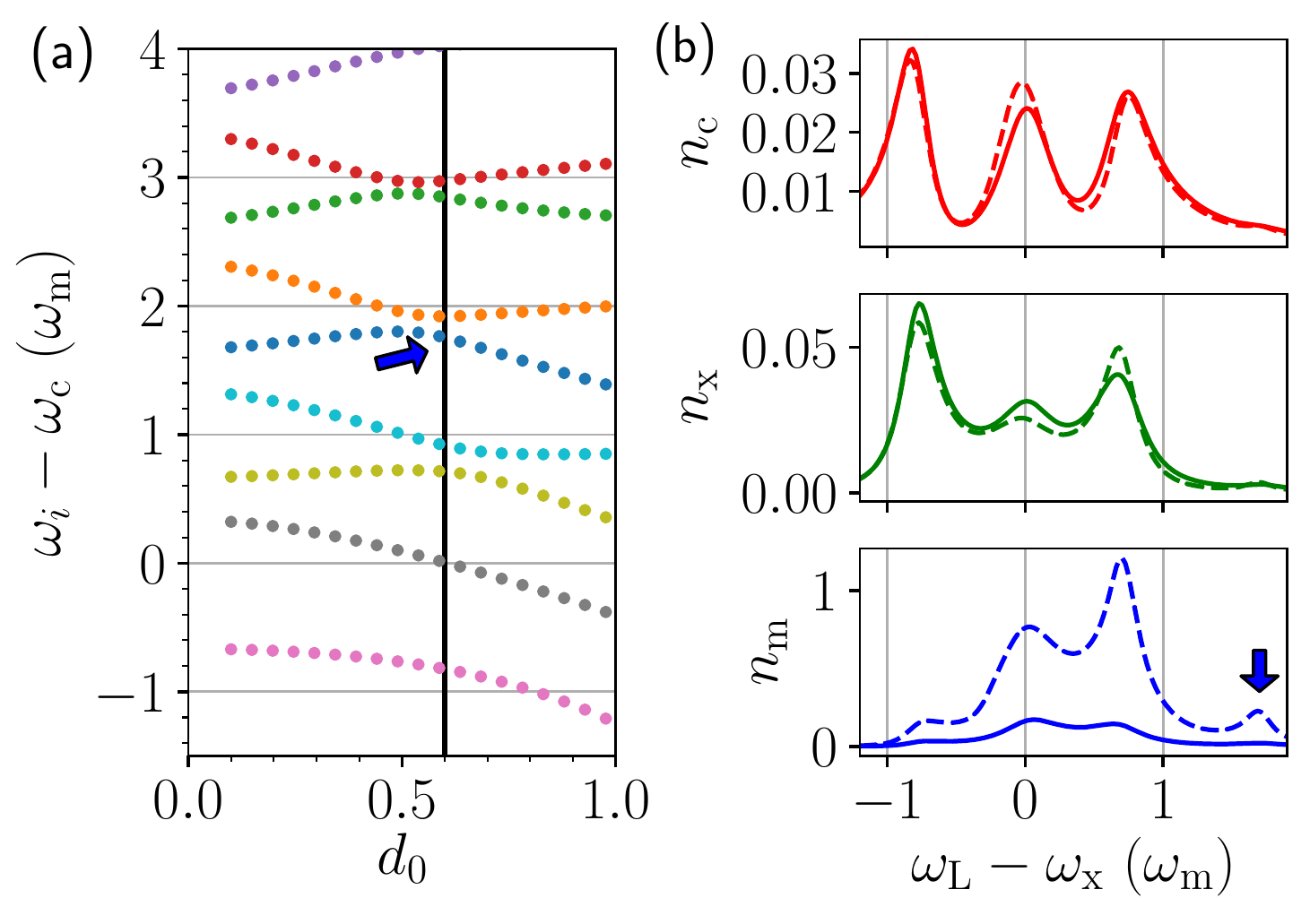}
\vspace{-0.5cm}
\caption{
(a) As in Fig.~\ref{fig:3}(a), but showing resonances
at $d_0=0.6$.
(b) Steady-state populations as a function
of laser detuning for the cavity, exciton,
and phonon.
The dashed (solid) curves
 show the SME (GME) solutions.
 \label{fig:4}}
\end{figure}

Finally, we  study how the dressed
resonances influence the 
 mean populations versus laser detuning, which can be accessed experimentally
from the photoluminescence
spectra~\cite{PhysRevB.86.241304}. 
The steady-state populations of  the cavity mode, vibrational mode, and TLS are obtained from from $n_{i}=\braket{x_{i}^-  x_{i}^+}$.
%
Figure \ref{fig:4}
shows the eigenenergies
at $d_0 =0.6$, using the same parameters
as in Fig.~\ref{fig:3}. Figure \ref{fig:4}(b)
displays the mean populations as a function of laser-exciton detuning, where the first phonon resonance near zero detuning is clearly showing up, and we also see the fifth resonance, which is the $k=1$ upper polariton state (marked by a blue arrow).
We note that the phonon populations are completely overstimated with the SME as the phonon displacement acts to increase the phonon damping in the vibrational USC regime. In addition, we see modified spectral asymmetries~\cite{Cao2011}
in both the cavity and exciton profiles.



\section{Conclusions}
In summary,  we have 
presented a GME approach 
to describe the regime of cavity-QED strong coupling under vibrational USC, demonstrating several rich regimes of single molecule optomechanics, in regimes where the SME significantly fails.  
We demonstrated how phonons {\it dress} the polariton states  of strong cavity-excion coupling,
which also leads to pronounced spectral asymmetries
from TLS pure dephasing and other bath-induced resonances,
even at low temperatures. With increased
exciton-phonon coupling, we also explored a regime
where both cavity coupling and phonon coupling
act in concert to produce new polariton states 
that appear below the Raman sidebands, and where fundamentally new resonances appear, which are visible in the cavity emitted and
photoluminescence spectra. In all cases, we showed explicitly
that the SME  significantly  fails.
All our predictions use system parameters similar to recent experiments and thus should be within
experimental reach.

\section*{Acknowledgements}

We acknowledge funding from 
the Canadian Foundation for Innovation and
the Natural Sciences and Engineering Research Council of Canada.
%
F.N. is supported in part by: 
Nippon Telegraph and Telephone Corporation (NTT) Research, 
the Japan Science and Technology Agency (JST) [via 
the Quantum Leap Flagship Program (Q-LEAP), 
the Moonshot R\&D Grant Number JPMJMS2061, and 
the Centers of Research Excellence in Science and Technology (CREST) Grant No. JPMJCR1676], 
the Japan Society for the Promotion of Science (JSPS) 
[via the Grants-in-Aid for Scientific Research (KAKENHI) Grant No. JP20H00134 and the 
JSPS–RFBR Grant No. JPJSBP120194828],
the Army Research Office (ARO) (Grant No. W911NF-18-1-0358),
the Asian Office of Aerospace Research and Development (AOARD) (via Grant No. FA2386-20-1-4069), and 
the Foundational Questions Institute Fund (FQXi) via Grant No. FQXi-IAF19-06.
S.S. acknowledges the Army Research Office (ARO)
(Grant No. 
W911NF-18-1-0358)
We thank Javier Aizpurua for useful comments and discussions.

\appendix*
\section{Connection between off-resonant SERS and resonant SERS from a polaronic picture}
\label{SI}

To make the connection between off-resonant SERS and
resonant SERS clear, 
it is  useful to consider the system Hamiltonians 
in a polaronic picture, which has
certain advantages in handling nonperturbative phonon (vibrational) coupling. Below we discuss how these two forms relate to each other, and when they are drastically different in general.
Similar analogies have been pointed out
in Refs.~\onlinecite{Dezfouli2019,PhysRevA.100.043422}, but without formal definitions of the polaron transforms.

Specifically, the resonant SERS scheme we study in the main paper is substantially different
to the off-resonant SERS scheme. The latter (off-resonanat form) is related to the usual optomechanical Hamiltonian, and the former (on-resonant form) fully recovers the physics of the Jaynes-Cumming with phonon interactions are turned off (and includes three system operators, not two). Our phonon interactions are also included without any approximations, as is required for studies in the vibronic ultrastrong coupling regime, requiring a full nonlinear and nonpertutbative treatment,
as well as a careful consideration for system-bath interactions.

 \subsection{Off-resonant Raman interactions} 

For  {\it off-resonant} Raman interactions (usual SERS form), we
 consider the standard optomechanical interaction without any form of linearization~\cite{RevModPhys.86.1391}, 
and neglect optical pumping,
\begin{align}
H_{ s}^{\rm off-res} & =\omega_c\, a^{\dagger}  a+\omega_{m}\,  b^{\dagger}  b+g_{\rm om}\, a^{\dagger}  a\left( b^{\dagger}+ b\right),
\label{eqA:H}
\end{align}
where  cavity operator terms $ a  a$ and $ a^{\dagger} a^{\dagger}$ (dynamical Casimir effects) can be safely ignored
here, as $\omega_c\gg \omega_m$.
The optomechanical coupling factor is given by~\cite{Roelli}
\begin{equation}
g_{\rm om}=\left( R_{m}/2\omega_{m}\right)^{-1/2}
\frac{\omega_{c}}{\epsilon_{0}V_{c}},
\end{equation}
 with $R_{m}$  the Raman activity\footnote{Note that the Raman activity is related to the elements of the Raman tensor, but they have different units.} associated with the vibrational mode under study, and $V_{c}$ as the effective mode volume of the cavity mode~\cite{Roelli}. Note the   same concept has also extended to arbitrary bath media~\cite{KamandarACS} (e.g., not just as the level of simple coupled mode theory).


Since the eigenstates are tensor products of fixed number states with displaced harmonic oscillator eigenstates, the polaron transform here is just a $ a^\dagger  a$-dependent displacement of the mechanical resonator:
$ \hat S=d_0 a^\dagger  a( b^\dagger-  b)$.
The polaron transformed system Hamiltonian for {\it off-resonant SERS},
$\tilde H_{\rm s}^{\rm off-res}
= e^{\hat S} H_{ s}^{\rm off-res} e^{-\hat S}$.
is then
\begin{align}
\tilde H_{ s}^{\rm off-res}
&=  (\omega_c-\Delta_{P})  a^{\dagger}  a +  \omega_{m}  b^{\dagger}  b
- \Delta_{P} a^\dagger  a^\dagger  a  a,
\label{eqA:H1}
\end{align}
where  $ a \rightarrow  a \hat X$, 
and the displacement operator is 
$\hat X=\exp[d( b-  b^\dagger)]$. In the polaron frame, the
effective cavity resonance is
shifted down by $\Delta_{P}=g_{\rm om}^2/\omega_m=d^2\omega_m$ (well known polaron shift), as also shown
from the eigenvalues of the optomechanical coupling problem (discussed in Sec~\ref{sec2} of our paper).
The Kerr-like term (4 operators) causes a nonlinear dependence on photon
number (photon-photon interaction), and {\it if} this term can be neglected, then one can write
\begin{align}
\tilde H_{s}^{\rm off-res}
\approx
 (\omega_c-\Delta_{P})  a^\dagger  a +
\omega_{m}  b^{\dagger} b,
%
\label{eqA:H2}
\end{align}
which is clearly in a much simpler form. Note the nonlinearities in this optomechanical coupling term
are not the same as a Fermionic system, since all operators here are bosons.

 \subsection{Resonant Raman interactions} 
 
For {\it resonant Raman interactions} (see Fig.~\ref{fig1}), 
the system Hamiltonian is now
\begin{align}
H_{s}^{\rm on-res}  &=  \omega_c  a^{\dagger}  a+
 \omega_{m} b  b^\dagger 
+  \omega_{x}  \sigma^+  \sigma^- + 
\nonumber \\
&+ d_0 \omega_m   \sigma^+  \sigma^-  \left( b^{\dagger}+ b\right) 
+ g( \sigma^+  a +   a^\dagger  \sigma^-),
\label{eqA:H3}
\end{align}
where the latter term is the Jaynes-Cummings term, and, as noted in the main text,  we  assume $g \ll \omega_c$,
so as to neglect USC effects related to the TLS-cavity coupling.
Note that we are now dealing with three system operators, two bosonic, and one Fermionic.
The polaron transformed system Hamiltonian, for resonant SERS,
now using $\hat S=d_0  \sigma^+  \sigma^-(b^\dagger-b)$
and $\hat X=\exp(d_0( b -  b^\dagger))$,
is
\begin{align}
\tilde H_{ s}^{\rm on-res}
&=   \omega_c a^{\dagger} a +
 \omega_{m} b  b^\dagger 
+  (\omega_{x}-\Delta_{P})  \sigma^+  \sigma^-
\nonumber \\
 &+  g( \sigma^+    a \hat  X +   a^\dagger \hat X^\dagger  \sigma^-).
\end{align}

The  equivalence between the off-resonant and resonant SERS can be made with a harmonic oscillator approximation for the TLS
in a bad-cavity limit.
Specifically,  the resonant SERS  polaron transformed Hamiltonian
then becomes identical in form to  the off-resonance case, apart from the
Kerr-like term (which vanishes for Fermions), and one simply
replaces
$ a, a^\dagger$ with $ \sigma^-,  \sigma^+$,
and identifies $d=d_0$.
When Fermionic behavior becomes important in the two state system
(e.g., Mollow physics, strong coupling between the
cavity mode and two-level system), then clearly
one must use the Pauli operators and the off-resonant SERS Hamiltonian is no longer appropriate.
Indeed, in a good cavity regime, and with nonlinear resonant excitation and cavity-QED interactions, the systems are vastly different.

\bibliography{Refs}

\end{document}